# Molecule Generation Experience: An Open Platform of Material Design for Public Users

Seiji Takeda[1*], Toshiyuki Hama[1], Hsiang-Han Hsu[1], Akihiro Kishimoto[1],
Makoto Kogoh[2], Takumi Hongo[2], Kumiko Fujieda[2], Hideaki Nakashika[2],
Dmitry Zubarev[3], Daniel P. Sanders[3], Jed W. Pitera[3],
Junta Fuchiwaki[4], Daiju Nakano[1]

[1] IBM Research – Tokyo, [2] IBM Garage, Tokyo Laboratory, [3] IBM Almaden Research Center, [4] JSR Corporation

{seijitkd, hama, hhhsu, dnakano, makotox, kfujideda, nakasika}@jp.ibm.com,
{dsand, pitera}@us.ibm.com, {akihiro.kishimoto, dmitry.zubarev, takumi.hongo}@ibm.com,
Junta_Fuchiwaki@jsr.co.jp

## ABSTRACT

Artificial Intelligence (AI)-driven material design has been attracting great attentions as a groundbreaking technology across a wide spectrum of industries. Molecular design is particularly important owing to its broad application domains and boundless creativity attributed to progresses in generative models. The recent maturity of molecular generative models has stimulated expectations for practical use among potential users, who are not necessarily familiar with coding or scripting, such as experimental engineers and students in chemical domains. However, most of the existing molecular generative models are Python libraries on GitHub, that are accessible for only IT-savvy users. To fill this gap, we newly developed a graphical user interface (GUI)-based web application of molecular generative models, "Molecule Generation Experience", that is open to the general public. This is the first web application of molecular generative models enabling users to work with built-in datasets to carry out molecular design. In this paper, we describe the background technology extended from our previous work. Our new online evaluation and structural filtering algorithms significantly improved the generation speed by 30 to 1,000 times with a wider structural variety, satisfying chemical stability and synthetic reality. We also describe in detail our Kubernetes-based scalable cloud architecture and user-oriented GUI that are necessary components to achieve a public service. Finally, we present actual use cases in industrial research to design new photoacid generators (PAGs) as well as release cases in educational events.

## 1. INTRODUCTION

Since the dawn of man, material has been a source of the power for the development of civilization equivalent to fire and electricity. The advent of semiconductors, liquid crystals, duralumin, conductive polymers, and other functional materials has drastically changed our lives across every industrial domain. Harnessing materials so as to bring desired characteristics (e.g. hard to burn, easy to melt, etc.) is a long-awaited technology that can revolutionize today's industries. However, material development to date has been driven by human expert's trial-end-error cycles conducted by using knowledge, experience, intuition, and serendipitous discoveries. Therefore, a typical lead time to design new material takes more than 10 years, and material's structures are most often limited by the individual creativity of each expert. Over the past decade, data-driven approaches leveraging artificial intelligence (AI) have been introduced to material science to increase design speed and the structural diversity. The four major technical domains to which AI and data have been introduced are, data collection and knowledge curation [1], acceleration of physical and chemical simulation [2], material's structural design [3], and autonomous synthesis [4]. Among those fields, structural design, especially molecular design, has been at the center of focus owing to its broad application domains and interesting problem definition on the context of generative models. For the sake of convenience, in this paper we call a class of AI-driven technologies that automatically generate brand new molecular structures as "*molecular generative models*", regardless of their background technologies; deep neural network (DNN), search algorithm, etc.

The primary requirement for molecular generative models is industrial practicality, including *high speed design*, *wide structural variety*, *chemical reality of designed molecules*, and *explainability and fine-tunability of a model*. Another important requirement is that molecular generative models should be accessible for a wide spectrum of users, who are most likely not data scientists. Core users are researchers, engineers, and students of experimental chemistry, all of whom are mostly unfamiliar with software-related works; installing appropriate packages, writing long scripts, using a command line interface, etc. Other users, considered to be more general and in higher numbers, include those who are not professionals but interested in, or in the early stages of learning AI and material science (e.g. scientific journalists, etc). Releasing a state-of-the-art technology to a wide variety of potential users over a barrier of coding, installing, etc., is important to facilitate agile and adoptive development for the dynamic evolution of science and technology.

## 2. RELATED WORK AND MOTIVATION

In this section, we first introduce state-of-the-art of molecular generative models and accessible public web services related to AI-driven material design, and we then identify unaddressed challenges that will be tackled in this paper.

### 2.1. Molecular Generative Models

Any organic materials (polymers, drugs, etc.) are molecule aggregates. A molecule has a graph structure whose nodes are atoms (carbon atom, etc.) and edges are chemical bonds (double bond, etc). The graph structure determines the properties of that molecule. The fundamental function required for molecular generative models is to design brand new molecular structures that can satisfy target property values required by a user (e.g. melting point should be higher than 150 °C). Several graph generative models have been reported.

***SMILES-based approaches***

A molecular graph can be represented by the Simplified Molecular Input Line Entry System (SMILES) strings. For example, caffein's molecular graph is represented as "CN1C=NC2=C1C(=O)N(C(=O)N2C)C". By leveraging matured natural language processing (NLP) with DNNs, many SMILES generation algorithms have been reported. A combination of a recurrent neural network (RNN) with a variational auto encoder (VAE) has paved the way in this field [5,6]. The grammar of SMILES is so fragile that the low success rate to generate valid SMILES has been a problem with a VAE. Therefore, several efforts have been spent, for example, implementing context-free grammar to the encoding and decoding schemes [7]. Improved sampling scheme in a latent space [8], and the integration of reinforcement learning (RL) [9] have resulted in more rapid and accurate molecule generation that are controlled by external conditions. Instead of using a VAE, a generative adversarial network (GAN) with an RL [10] and a Monte Carlo Tree Search (MCTS) with an RNN have been also proposed [11].

***Graph-based approaches***

While the SMILES-based approaches have been gaining significant interest, there has been another stream of approaches in which a molecule is encoded by raw-graph representations. In MolGAN, an adjacency tensor and annotation matrix are used to represent a molecule in combination with a GAN and an RL [12]. In Junction-Tree VAE, a molecule is represented by a molecular graph and a tree. Each node equals a substructure and a molecule is generated by connecting those nodes [13]. In the Hyper Graph Grammar method [14], a molecular graph grammar is generated using a VAE, leading to the improvement of grammar accuracy.

Our recent approach has exploited a graph enumeration algorithm for molecular generation in combination with regression and optimization algorithms [3]. Molecules generated by SMILES-based generative models are influenced by the domain of pre-training dataset, however in practice, chemists do not always own a large amount of training data in their advanced material domains. Additionally, they need to understand why molecules newly generated are reasonable. Our approach consists of an encoder and a decoder defined by algorithms. It is advantageous that pre-training using a big data is not required and the model is fully explainable from chemical viewpoints and as well as fine-controllable at the atomistic level.

### 2.2. Public Services for Material Design

While there are many libraries related to material design available on GitHub for software developers and cheminformatists, there are only few graphical user interface (GUI)-based web applications accessible to non-IT-skillful users exist. Polymer Genome is a web application on which a user can predict polymer's properties [15]. IBM RXN for Chemistry provides a variety of functions related to chemical reactions [16]. PaccMann enables a user to predict an anticancer compound's sensitivity to a target [17]. Potential drug candidates for COVID-19 can be browsed in [18]. Those web applications provide functions of property prediction or reaction prediction on a pre-trained model. However, to the best of our knowledge, there have been no web applications for molecular generative models that are open to the general public.

### 2.3. Challenges to Address

Existing molecular generative models, especially SMILES-based DNN ones, have achieved landmark success to bridge NLP with the field of molecular chemistry. However, from a viewpoint of industrial practicality, unsolved challenges still exist. Due to the nature of DNNs, models are hard to explain, and fine-controllable molecular generation is difficult because new candidate molecules are stochastically sampled from a latent space, which has no detailed information in terms of molecular structures in atomistic granularity. In our previous work [3] we have addressed those issues with making the feature space and generation process fully explainable for chemists and molecular generation fine-controllable by atomistic constraints. Nevertheless, we still have identified additional challenges that need to be addressed for industrial use, including (1) the speed of generation, (2) the variety of generated structures, and (3) the chemical reality of generated structures. In our previous approach and most DNN approaches, the generation speed has been limited by the process of samplings from a feature (or latent) space. A sampled feature vector obtained by solving an optimization problem on the targe property, with reference to a trained regression model, is typically infeasible to be decoded into a molecule, because decodable features are distributed only on a tiny manifold in the feature space. The manifold is rigorously defined by implicit rules, so it is unclear whether a molecular structure can be successfully generated from the sampled feature vector until the generation process is run. In this paper, we overcome this challenge by eliminating optimization and sampling processes, and instead developing online evaluation algorithms. Another important issue is that (4) there are no molecular generative models available in the form of a web application, so the technology is still far from potential public users' reach. Addressing the above issues not only enables a practical molecular generative model service, but also lays out a new style of an open science platform in AI and chemistry.

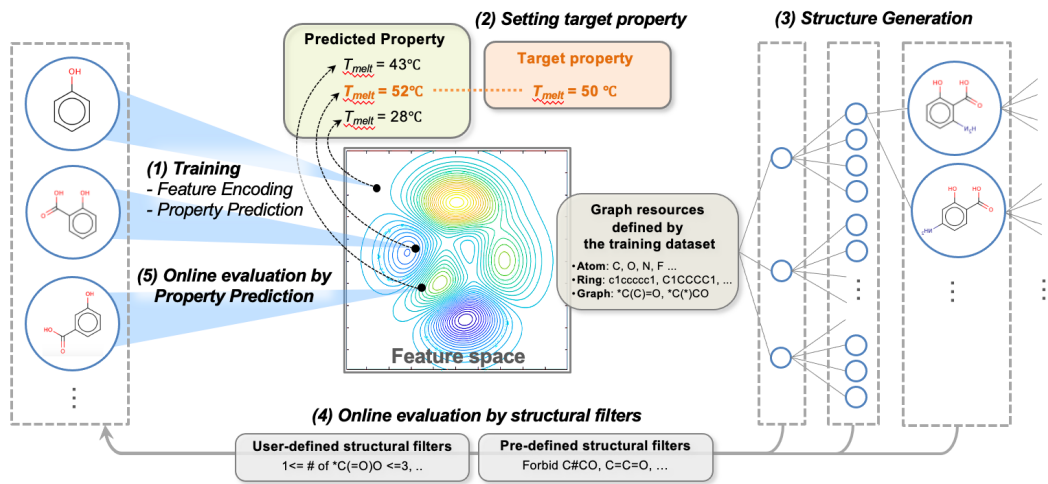

**Figure 1: Fundamental workflow of Molecule Generation Experience.**

## 2.4 Contributions

Given the aforementioned unsolved issues, in addition to the foundation of our latest tool, we developed a new graph generation algorithm. We present the following contributions.

- Developing an online property prediction method, that significantly improved the speed of structure generation and the variety of structures generated.

- Developing an online structure-filtering algorithm that adopts structural rules during the generation process to efficiently screen out undesirable structures.

- Developing a list of chemical rule-based structural filters that is implemented in the filtering algorithms.

- Developing a Kubernetes-based cloud architecture on which the API set is orchestrated in a scalable manner.

- Developing a new GUI that summarizes the complex molecular generation workflow to three simple steps.

- Releasing the above tool set to the general public as "Molecule Generation Experience (MolGX)".

## 3. METHODS

The distinct difference between MolGX and other molecular generative models is that the encoder and decoder are configured by algorithms. Additionally, in this paper we present our new online prediction and filtering methods.

### 3.1 Overview

The fundamental workflow of MolGX is illustrated in Figure 1. The workflow is composed of five steps. First, a user (1) trains a property prediction model with feature encoding and property prediction by inputting a material dataset. Next, the user (2) sets a target property, and then the system carries out (3) structure generation, that is evaluated in terms of (4) structural filters and the (5) target property. Steps (3-5) are iterated until resources for generation are depleted. Detailed descriptions are given as follows.

### 3.2 Feature Encoding and Property Prediction

For this part, we apply our previous work [3], and summarize the key steps. To encode a molecular structure as a feature vector, we have exploited an approach based on a graph kernel concept. MolGX encodes each molecular graph by the frequencies of substructures appearing in it. MolGX then classifies the features, referring to the substructures into several types having different granularities; {*atoms*}, {*rings*}, {*aromatic rings*}, {*substructures having one edge, two edges, .. *} etc., from which a user can select arbitrary combinations to use for encoding. This encoding scheme is advantageous in its complete explainability for chemists. Its property prediction accuracy is as high as other encoding schemes (e.g. Morgan fingerprint [19]). MolGX finally encodes an input data set, which is given as pairs of molecular structures and associated properties, into a set of feature vectors in accordance with the above scheme.

After the feature encoding process, MolGX trains a regression model to predict a target property. The user can manually change the hyper-parameters, but MolGX automatically optimizes all model configurations including a combination of features, a type of model, and hyper-parameters of each model.

## 3.3  Structure Generation with Online Evaluation

As described in Section 2.3, the generation speed and structural variety are mainly limited by the sampling process in a feature space, where most of the sampled feature vectors are infeasible to be decoded. In this paper, we shift away from the optimization and sampling processes. In our new approach, we generate a variety of molecular structures in parallel with their evaluations in terms of structural rules and target properties. At first glance, this concept might look similar to virtual screening. However, our approach has the following notable difference. In virtual screening, a pre-built exhaustively large library of molecular structures, including a significant number of uninteresting structures, is screened by prediction models and other structural rule sets as a post hoc process, resulting in substantial wasteful generation and prediction processes. In contrast, our online evaluation algorithm prunes unnecessary graph generation paths during the early stages of generation, so generates only desired molecules meeting the criteria. We summarize the workflow in Algorithm 1-3. Overlapped parts with our previous work especially in Algorithm 2 are simplified, so please refer to [3] for more detail. The following subsections describe each process.

### Graph-building process

We have improved the graph-building approach, Molecular Customized Mckay's Canonical Construction Path (MC-MCCP) [3] to integrate the online evaluation. In MC-MCCP, molecular graphs are built by repeatedly connecting *graph resources*, which consists of atoms, rings, and user-defined substructures, while avoiding isomorphic duplications by computing a canonical label. The range of graph resources, that is, the number of atoms, rings, and substructures to be "consumed" in the building process, is set to be identical to that of the original dataset. The process produces a trace of a tree-shaped generation path (see Figure 1) by repeating iterations of the following steps. (i) A graph resource is connected to a possible position on a current molecule, (ii) a canonical label is computed to check duplicability, (iii) the generated structure is evaluated by structural filters and (iv) the structure's property is also evaluated. In our previous method, the third step

---

**Algorithm 1: Main algorithm**

**Input:** #Atoms // dictionary of atoms and numbers
    #Fragments // dictionary of fragments and ranges
     R // structural rules
     M // regression models
**Output:** molecular structures as graphs
1: **generate**(#Atoms, #Fragments, R, M)
2:   **Foreach** a **in** #Atoms.keys
3:     g = <{a}, {}> /* new graph */
4:     #Atoms0 = *decrement* #Atoms[a]
5:     *augment*(g, #Atoms0, #Fragments, R, M)
6:   **End**

**Algorithm 2: Graph-building**

1: **augment**(g, #Atoms, #Fragments, R, M)
2:   **If** *evaluate_structure*(g, #Fragments, R, M)
3:     *output* g /* solution found */
4:   **Endif**
5:   **If** *check_termination*(g, #Aoms, #Fragments)
6:     **return** /* no possibility in future */
7:   **Endif**
8:   /* augment a graph by adding a new atom */
9:   /* iterate valid combination of vertex, atom, bond */
10:   **Foreach** v, a, b **in** *canonical_comb* (g, #Atoms, {1,2,3})
11:     e = *edge*(v, a, b)
12:     g0 = <g.V+{a}, g.E+{e}>
13:     #Atoms0 = *decrement* #Atoms[a]
14:     *augment*(g0, #Atoms0, #Fragments, R, M)
15: **End**

**Algorithm 3 : Online evaluation**

1: **evaluate_structure**(g, #Fragments, Rules, Models)
2:   /* check acceptable range of fragments */
3:   **Foreach** fragment **in** #Fragments.keys
4:     range = #Fragments[fragment]
5:     **If** not range.*include*(*count_fragment*(g, fragment))
6:       **return** False
7:     **Endif**
8:   **End**
9:   /* check structural rule */
10:   **Foreach** rule **in** Rules
11:     **If** not rule.*accept*(g)
12:       **return** False
13:     **Endif**
14:   **End**
15:   /* check prediction value */
16:   **Foreach** model **in** Models
17:     **If** not model.*in_target_range*(g)
18:       **return** False
19:     **Endif**
20:   **End**

---

checked the generated molecule's feature with the sampled feature point. In this paper, to further enhance efficiency as well as to improve a variety of generated molecules, we replace that step with the aforementioned two steps, which are described in more detail in the following subsections.

**Table 1: Examples of pre-defined substructures excluded in generated molecules.**

| Substructure | Name | Reason |
|---|---|---|
| C=C≡N | ketenimine | Unstable |
| C#CO | ynolate | Unstable |
| C=C=CO | allenolate | Difficult to synthesize |
| ONO | N,N-dihydroxylamine | Difficult to synthesize |
| ONC=C | enamine-N-oxide | Difficult to synthesize |

***Online structural filtering***

At each step of generation, our improved MC-MCCP evaluates a new molecular graph generated by adding a graph resource. In this step, based on the structural rules pre-defined by a user and the system's default setting, our approach ensures a structural desirability and its chemical reality on the molecule graph under construction. More specifically, the structure is checked with each structural rule; for example, a range of the number of specific substructures or structural templates to be included in a molecule, chemical bonding type to be satisfied, etc. If the structure satisfies the rule, it is passed to the next process (iv) online property prediction. Otherwise, the current generation path is pruned, and the process switches to another path on which a new generation starts. On top of the canonical label check in (ii), our new pruning process significantly suppresses generation processes on unnecessary paths.

***Online property prediction***

Instead of solving an optimization problem of feature vectors on a target property, MolGX directly evaluates a generated molecule's property, using the trained regression model at each generation step of the graph building algorithm. The molecular structure that has passed the structural filtering is encoded into a feature vector and then evaluated on the regression model. If the predicted property satisfies the user-set target value (e.g. $E_{lumo} \sim -2.5$ eV), the structure is accepted and stored in the candidate list. Regardless of whether the new structure satisfies the target property or not, the iteration from (i) to (iv) continues until all the graph resources are exhausted or the current generation path is pruned. This iteration enables to generate a much wider variety of molecular structures than our previous approach.

## 3.4 Chemical Filters

We carry out a structural screening by a set of rules on substructures, chemical bonds, etc. A user can arbitrarily set a rule, but in MolGX, we have implemented a default rule set that is pre-defined by chemists. The list consists of ∼ 50 chemical structural rules. Each rule defines an unrealistic substructure that should be excluded in chemical compounds from a viewpoint of chemical stability and difficulty to synthesize. Example rules are shown in Table 1. As will be described in a later section, this filtering reduces a considerable number of unrealistic structures. This rule set describes only the minimum rules that are common to general chemical domains. In practice, users need to define their own rules that are useful in a specific material domain.

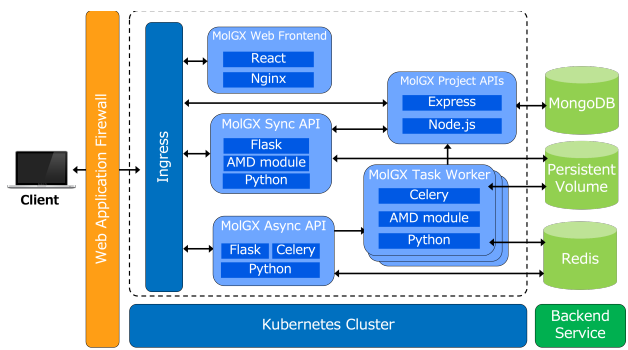

**Figure 2: Kubernetes-based cloud architecture of**

## 4. ARCHITECTURE OF WEB APPLICATION

We have developed the aforementioned algorithms as a Python package. To configure it as a GUI-based web application for a wide spectrum of users, we have designed a workflow from a viewpoint of user experience (UX), developed API set, and implemented it on a Kubernetes-based cloud environment, that works on IBM Cloud.

### 4.1 Workflow Design

To develop a web application, it is crucial to provide a comfortable UX. This involves designing a user-friendly workflow, which should be easy to understand the whole steps and each step the user needs to follow. We streamline the MolGX's complex workflow to three sequential steps. (1) *Process dataset*, where a user observes the distribution of data and selects a subset that is used to train a model; (2) *Train AI*, where the user selects features to use, and trains the model to predict a target property; and (3) *Design molecules*, where the user sets a target property value and starts molecular generation. The user can intuitively navigate between those steps switching datasets, models, and hyper-parameters to obtain desired results.

### 4.2 API Configuration

Unlike general web applications for browsing contents or carrying out simple predictions, a molecular generative model requires long processing times, ranging from a few seconds to several minutes especially in model training and structure generation processes, which exceeds typical connection times. Therefore, we have implemented these time-consuming processes as asynchronous APIs, and other processes (e.g. displaying results) as synchronous APIs. Table 2 shows an example of the configured REST API. By combining these API sets, MolGX efficiently allocates processes to backend or frontend servers depending on their CPU loads.

**Table 2: Examples of Synchronous and Asynchronous**

| Method | API Path | Params | Response (JSON) | Type |
|--------|----------|--------|-----------------|------|
| PUT | /api/v1/projects | Project Data | Update Result | Synchronous |
| GET | /api/v1/designResults/:projectId | Project ID | Design Result | Synchronous |
| POST | /api/v1/tasks | Task Data | Submission | Asynchronous |
| GET | /api/v1/tasks/:tas | Task ID | Task | Asynchronous |

## 4.3 Cloud Architecture

In our previous work, we implemented our tool onto a virtual machine (VM) in a local environment, because of the limited user accesses to the tool. However, a public web application needs scalability in terms of the number of users who can access the system simultaneously. Additionally, MolGX runs various machine learning tasks on the basis of user requests. Multi-node clustering is necessary for better performance, but it is difficult to configure it on conventional VMs. To improve portability, high availability and scalability, MolGX adopts a micro-service architecture on Kubernetes. By leveraging Kubernetes, processes are automatically distributed across multiple worker nodes, so that parallel processing is dynamically configured in accordance with the number of accessing users and their processes. To take advantage of Kubernetes, we have developed the architecture by following "Beyond the Twelve-Factor App" rules. The architecture is presented in Figure 2.

## 5. PERFORMANCE EVALUATION

We run our end-to-end workflow to design new molecular structures satisfying a targeted property and evaluate the performance in terms of speed, variety, and chemical reality. We describe the details in comparison with our previous work.

## 5.1 Data Preparation and Property Prediction

We used the QM9 dataset, which is one of the typical benchmarking sets in this domain [20]. QM9 includes 134,000 small organic molecules with 17 associated chemical properties calculated by density functional theory (DFT) simulation. From the original set, we extracted 300 different molecules for the training and test sets, respectively to make the properties' distributions similar between the sets (see Figure 3). These sets are consistent with the actual web application.

We trained a property prediction model on the energy of the highest occupied molecular orbital (HOMO); $E_{HOMO}$. We configured a molecule's feature types by numbers of {*heavy atoms*, *rings*, *aromatic rings*, *substructures including one edge*}, resulting in 106-dimensional feature vectors. We trained four regression models (lasso, ridge, kernel ridge, support vector) by sweeping the hyper-parameters for each model. By comparing the average validation scores using 10-fold cross validation, we selected the model kernel ridge, which produced reasonable accuracy ($R^2_{train} \sim 0.75$, $R^2_{test} \sim 0.68$) considering the small size of the dataset. We use this model for the following molecular generation process.

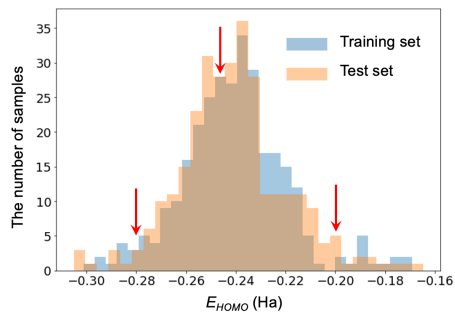

**Figure 3: $E_{HOMO}$ distribution in training and test datasets. Red arrows represent the three target values for structure generation.**

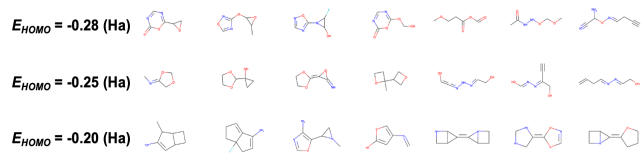

**Figure 4: Examples of molecular structures generated by MolGX targeting each $E_{HOMO}$.**

## 5.2 Speed of Generation

In structure generation, we used three target properties; $E_{HOMO} \sim -0.28$ (Ha), $-0.25$ (Ha), and $-0.20$ (Ha) for the evaluation because the generation process is influenced by the density of the original data around the target value. We carried out generations by using our previous tool denoted as Molecular Inverse-Design Platform (MIDP) [3] and our new MolGX under the identical experimental conditions (data, prediction accuracy, and target properties). The speed was measured for the generation process. MIDP generated structures starting with 20 feature vectors that were already sampled out by solving optimization problem on a regression model. MolGX performed the generation process with online prediction and without the sampling process.

Running the tools until graph resources were consumed, we obtained 41 to 3,362 molecules in 21 to 50 min by MIDP, and 11,464 to 14,898 molecules in 6 to 10 min by MolGX. The measurement results are shown in Table 3. Depending on the target $E_{HOMO}$ value, the generation speed (i.e. the number of molecules generated per second) ranges from 0.03 to 1.1 sec$^{-1}$ for MIDP and 29 to 77 sec$^{-1}$ for MolGX. We confirm significant improvement; 33, 70, and 1,193 times faster for each target $E_{HOMO}$, respectively. The speed of other state-of-the-art generative models is typically 1 to 10 sec$^{-1}$ when QM9 is used [12]. This indicates that MolGX is competitive to other algorithms. We observe the speed is at maximum for $E_{HOMO} = -0.25$ (Ha) where the data is most densely distributed.

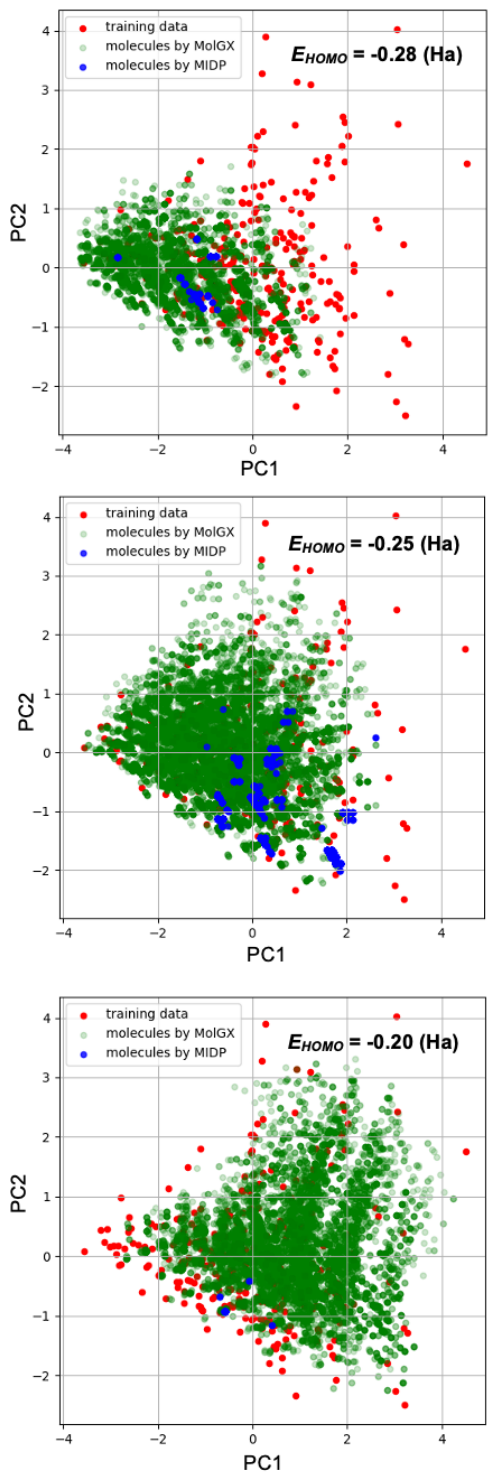

**Figure 5: PCA components of feature vectors of molecular structures that are included in the training dataset, those generated by MolGX in 6 to 10 min, and those generated by MIDP in 21 to 50 min. Each panel represents different target HOMO energy.**

**Table 3: Comparison of the number of generated molecules per single second targeting different $E_{HOMO}$.**

| Method | $E_{HOMO}$ = -0.28 | $E_{HOMO}$ = -0.25 | $E_{HOMO}$ = -0.20 |
|--------|------------------|------------------|------------------|
| **MolGX** | **28.9 (sec⁻¹)** | **76.8 (sec⁻¹)** | **38.5 (sec⁻¹)** |
| MIDP | 0.873 (sec⁻) | 1.10 (sec⁻¹) | 0.0323 (sec⁻ |

**Table 4: Comparison of the number of chemically valid or invalid molecular structures generated by applying or not applying structural filters.**

| Filter | $E_{HOMO}$ = -0.28 | | $E_{HOMO}$ = -0.25 | | $E_{HOMO}$ = -0.20 | |
|--------|-------|---------|-------|---------|-------|---------|
| | Valid | Invalid | Valid | Invalid | Valid | Invalid |
| **Filter-on** | **9,826** | 0 | **40,036** | 0 | **13,960** | 0 |
| Filter-off | 8,059 | 3,405 | 38,543 | 7,992 | 13,672 | 1,226 |

## 5.3 Variety of Generated Structures

A portion of generated structures are shown in Figure 4, where we confirm considerable varieties. To evaluate the variety of those generated structures, for each target $E_{HOMO}$, we mapped the principal component analysis (PCA) components of feature vectors corresponding to the molecules of the training dataset, those generated by MolGX, and those generated by MIDP (See Figure 5). Clearly MIDP tends to sparsely distribute its blue plots around a small area. In contrast, MolGX widely spreads its green plots over a larger area with a higher density even in the shorter generation period. Some portions of the structures generated by MolGX are distributed outside the boundary of the training dataset. This indicates the possibility that MolGX has an advantage of generating molecules belonging to a family of new structural patterns different from the existing molecules. On the other hand, we can also see that the entire range of the training dataset is not covered in any target property. By setting a graph resource's range wider than that of the training dataset, it is possible to expand the structural variety even wider, however, this is an extrapolation that reduces the model's credibility and is not currently addressed in this paper.

## 5.4 Chemical reality

Finally, we evaluate the effect of the chemical filtering process on the generated structures' quality. To eliminate the effects of other improved functions and focus on only the filter defined in 3.4, we compare the number of chemically-realistic (valid) structures obtained by running MolGX with/without applying the filter (hereafter, on-filter and off-filter, respectively). For on-filter cases, generated structures are naturally valid structures. For off-filter cases, realistic structures are obtained by screening all the molecules after they are generated. The results are shown in Table 4. Most importantly, despite the structures being screened, we observe that the number of valid structures in on-filter cases is higher than that of off-filter cases. This is because the structures are screened in parallel with generation, so unpromising generation paths are pruned as explained in Section 3.3. This path pruning process enables the system to explore other possible structures (*leaf nodes* of the generation path tree) with the limited resources (maximum number of leaves to explore). This is MolGX's clear advantage confirmed for on-filter cases.

## 6. WEB APPLICATION IN SREVICE

We have deployed all the components with a GUI as a web application service. We have started running the system for trial use by internal technical and non-technical communities, and then launch it online to the general public on https://molgx.draco.res.ibm.com (*MolGX is already deployed on the cloud, but the firewall will be opened at the end of February due to security policy). Users can login to MolGX using a Twitter, Facebook, or Google account. In this section, we will describe the details of the interface and introduce our actual industrial and educational use cases.

### 6.1 Graphical User Interface

Following the workflow described in Section 4.1, we designed a GUI that provides intuitive operations like clicking a button as well as visually guiding a user to the next action in the sequential workflow. MolGX has three main panels (see Figure 5), each of which corresponds to the process described in Section 4.1. Here, we describe the functions provided in each process.

*Process data*
- **Select a dataset** from three types of built-in datasets; QM9, Drugs, and photoacid generators. Each dataset includes 300 training samples and 300 test samples.
- **Select properties** of X/Y axes for the scatter plot.
- **Select samples** that need eliminated from the training set.

*Train AI*
- **Select feature encoding schemes** to encode each molecule to a feature vector before training.
- **Select target property** to predict by a regression model.
- **Select a regression model** used for training.
- **Set generalization level**, which is the degree of regularization. For simplicity, only a single controllable parameter is available in each regression model (e.g. kernel width), and the other parameters are fixed.

*Design molecule*
- **Set target property value** that needs to be satisfied by structures generated.
- **Add structural constraints** that are given by a substructure described in a SMILES expression and the minimum and maximum numbers to be included in the generated structures.
- **Set number to generate,** which is the number of molecular structures a user wants to generate.

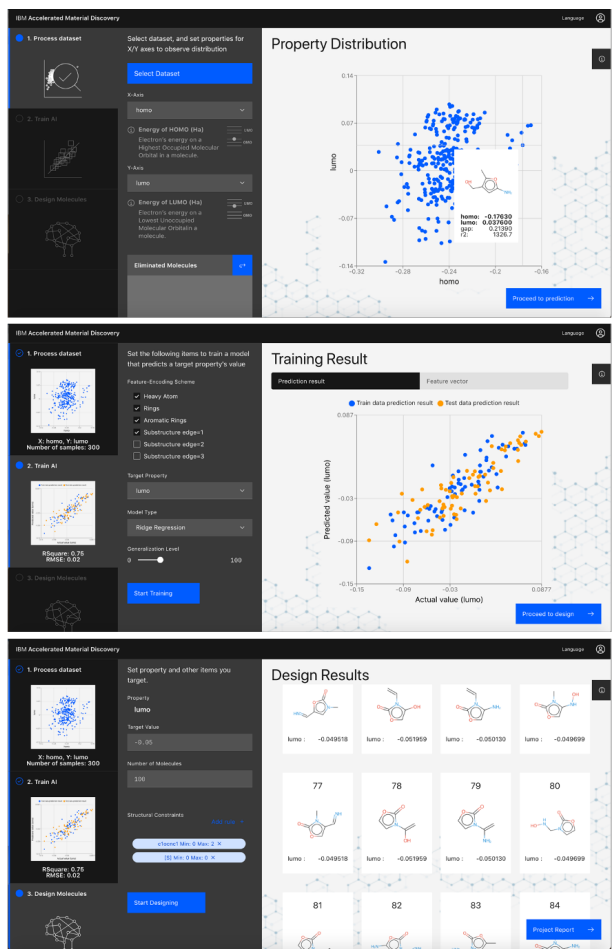

**Figure 5: Screen shots of MolGX, each corresponding to Process Data, Train AI, and Design Molecules.**

user can navigate between those panels to change a dataset, re-train a model using other parameter sets, etc. In addition to the above panels, MolGX has a report page on which a user can see the summary of used parameters, and interact with a generated molecule in a 3D virtual space.

### 6.2 Use Case Examples

Prior to public release, we carried out a limited release of MolGX for actual material industries and researchers, where a significant increase in material design speed was confirmed. For industrial cases, we included additional functions excluded in the public service; targeting multiple properties, setting detailed structural constraints on chemical bonds, etc. We also carried out a release in educational events where broad range of non-IT-professional users could access MolGX.

*Photoacid Generator*

We applied MolGX to an R&D project, in which we designed new molecular structures of photoacid generators (PAGs), that belong to sulfonium family [21]. A PAG is a vital material that plays a central role in today's semiconductor lithography. A PAG in a photoresist absorbs ultraviolet light and generates a strong acid, which in turn catalyzes chemical reactions within the resist material in the exposed regions, thereby enabling the creation of fine patterns used in the fabrication of semiconductor devices. With the rapidly increasing presence of semiconducting technology in our lives, we focus on proactive discovery of materials with minimized environmental impact that are consistent with evolving regulations. Simultaneously, we pursue further improvement of PAGs functionality in lithographic applications, where photosensitivity plays critical role. We trained our model with ~1,300 PAG structures extracted from U.S. patents, and corresponding property data which we calculated by DFT simulation. We targeted five properties; LogWS (solubility in water), LogP (lipophilcity/hydrophobicity), $T_{bio}$ (bio-degradation time), LD50 (toxicity), and $\lambda_{abs\_max}$ (peak wavelength of light absorption), satisfying specific ranges set by a PAG chemist. After running MolGX for 6 hours, more than 2,000 molecular structures were generated (see Figure 6). They were down-selected by a screening model trained by a human expert, and their synthetic accessibility was confirmed by using RoboRXN [16]. The reason for the low speed generation (~0.1 sec⁻¹) compared with the experiments in Section 5.2 is that PAGs have more complicated structures, with the number of heavy atoms being more than 20, while QM9 uses only 9. Also, we targeted five properties instead of one. Despite that, the design speed is amazingly high—more than several tens of times faster compared with a human expert.

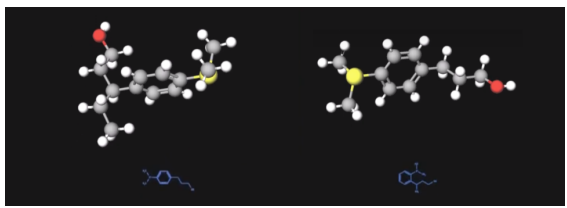

**Figure 6: Examples of generated PAG structures.**

***Educational events***

During several internal educational events (e.g. education sessions for sales teams, events for employees' families including elementary school students), we provided access to MolGX to pre-registered users. In each event, MolGX was accessed by more than 100 total users in which ~80% were not experts. MolGX marked high net promoter score (NPS); 87 to 91, while average scores of those events were 78 to 85, confirming MolGX's high accessibility and capability to penetrate the general public society.

## 7. CONCLUSION

We introduced the first public web application of molecular generative model, "Molecule Generation Experience (MolGX)".

The platform provides an experience of simplified end-to-end workflow to design new molecular structures by running our molecular generative model on built-in data sets. In this paper, we presented our new online evaluation and structural filtering algorithms that significantly improve generation speed, structural variety, and chemical reality, that were achieved by newly developed online evaluation and structural filtering algorithms. Comparing with the previous work, we demonstrated that generation speed was accelerated 30 to 1,000 times (generates ~76 molecules per second in maximum) with expanded variety of molecular structures, satisfying chemical reality. We also constructed Kubernetes-based cloud architecture which realizes scalability to handle multiple user access, as well as implemented user-oriented GUI. As actual use cases, we exhibited design of new PAG structures having low toxicity with specific optical characteristics, where several 10 times of acceleration in design speed comparing with human experts was confirmed. Finally, we introduced more educational and illuminating use cases at educational events for novice science learners. MolGX will serve as a user-friendly scientific platform for both professional and general public communities.

## ACKNOWLEDGMENTS

We thank to Dr. Peter W J Staar and Dr. Teodoro Lainio of IBM Research – Zurich, and Dr. Edward Pyzer-Knapp of IBM Research UK for their collaboration in the photoresist projects.